\documentclass[preprint,12pt]{elsarticle}

\usepackage{amssymb}
\usepackage{amsmath}

\usepackage{subfigure}
\usepackage{slashed}

\newcommand\nn{\nonumber}
\newcommand\ba{\begin{eqnarray}}
\newcommand\ea{\end{eqnarray}}
\newcommand\eq[1] {\begin{align} #1 \end{align}}   
\newcommand\ga[1] {\begin{gather} #1 \end{gather}}   

\newcommand{\br}[1]{\left( #1 \right)}
\newcommand{\brs}[1]{\left[ #1 \right]}
\newcommand{\brf}[1]{\left\{ #1 \right\}}
\newcommand{\brm}[1]{\left| #1 \right|}

\newcommand{\bra}[1]{\left< #1 \right|}
\newcommand{\ket}[1]{\left| #1 \right>}

\newcommand{\Sp}{\mbox{Tr}}

\renewcommand{\Re}{\mbox{Re}}
\renewcommand{\Im}{\mbox{Im}}

\newcommand{\vv}[1]{{\bf #1}}

\newcommand{\M} {{\cal M}} 
\newcommand{\dd}[1]{{\slashed #1}}   

\newcommand{\GeV}{\mbox{GeV}}
\newcommand{\MeV}{\mbox{MeV}}
\newcommand{\keV}{\mbox{keV}}

\begin{document}

\begin{frontmatter}



\title{About the creation of proton--antiproton pair at electron--positron collider in the energy range of $\psi(3770)$ mass}


\author[JINR,AZER]{A.~I.~Ahmadov}
\ead{ahmadov@theor.jinr.ru}

\author[JINR]{Yu.~M.~Bystritskiy}
\ead{bystr@theor.jinr.ru}

\author[JINR]{E.~A.~Kuraev}

\author[CHINA]{P.~Wang}
\ead{wangp@ihep.ac.cn}

\address[JINR]{Joint Institute for Nuclear Research, 141980 Dubna, Moscow Region, Russia}
\address[AZER]{Permanent address: Institute of Physics, Azerbaijan National Academy of Science, Baku, Azerbaijan}
\address[CHINA]{Institute of High Energy Physic, Chinese Academy of Science}

\begin{abstract}
    The process of electron--positron annihilation into proton--antiproton pair is considered within the vicinity of
    $\psi(3770)$ resonance. The interference between the pure electromagnetic intermediate state and the $\psi(3770)$ state
    is evaluated. It is shown that this interference is destructive and the relative phase between these two contributions is
    large ($\phi_0 \approx 250^o$).
\end{abstract}


\end{frontmatter}


\section{Introduction}

Large statistics of $J/\psi$, $\psi(2S)$ and $\psi(3770)$ samples have been obtained in recent years by BEPCII/BESIII
facility \cite{Ablikim:2012dx}.
It provides the possibility to study many decay channels of $J/\psi$, $\psi(2S)$ and $\psi(3770)$ resonances.
%
In a profound work, BESIII has measured the phase angle $\phi$ between the continuum and
resonant amplitudes \cite{Ablikim:2014jrz} and found two possible solutions, which are
$\phi = \br{266.9 \pm 6.1 \pm 0.9}^o$ or $\phi = \br{255.8 \pm 37.9 \pm 4.8}^o$.
This means that the strong decay amplitude and electromagnetic decay amplitude are almost
orthogonal.
The BES III data were taken as an energy scan in the vicinity of $\psi(3770)$. The data show some structure:
clearly seen dip in the energy strip of size of the resonance
$\psi(3770)$ width, which had been observed previously by CLEO collaboration for some mesonic
decay channels \cite{Adams:2005ks}.

In this note we try to explain this rather specific behavior of the total cross section of process $e^+e^- \to p\bar{p}$ in the energy range
close to resonance $\psi(3770)$ creation.

In contrast to the channel $e^+ e^- \to \psi(3770) \to \mu^+\mu^-$, in process of hadron creation
(i.e. $e^+ e^- \to \psi(3770) \to \pi^+\pi^-, \bar p p, \bar n n$), a QCD gluonic state contribution to the hadron
(in particular nucleon) formfactor $\psi \to \bar p p$ is to be investigated.
Besides the Breit--Wigner character of the amplitude, one must take into account the specific character of interaction of quarkonium to
nucleon--antinucleon pair mediated through 3 gluon intermediate state and the final state interaction of the created nucleon pair.

The second effect is the final state interaction phase of amplitude which arise mostly from large distances (or soft exchanges of final stable hadrons). It has the same form for
$\gamma^* \to \bar p p$ and for $\psi \to \bar p p$ vertexes and we can safely assume its cancellation
in the interference of pure QED and quarkonium states.

On the contrary, the phase which arises from 3 gluon state can essentially affect on the Breit--Wigner character of pure QED final state.

It is the motivation of this paper to investigate the detailed behavior of the total cross section in the energy range within the mass of a
narrow resonance $\psi(3770)$.

\section{Born approximation}

\begin{figure}
    \centering
    \mbox{
        \subfigure[]{\includegraphics[width=0.4\textwidth]{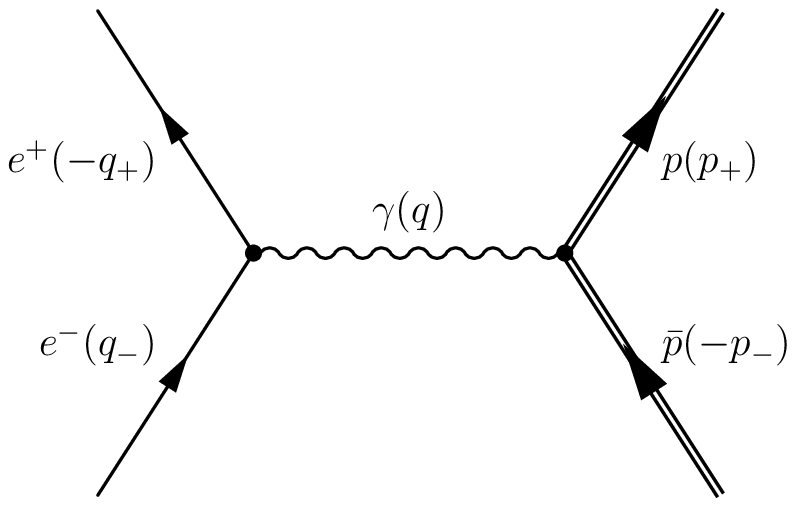}\label{fig.BornDiagram}}
        \quad\quad
        \subfigure[]{\includegraphics[width=0.4\textwidth]{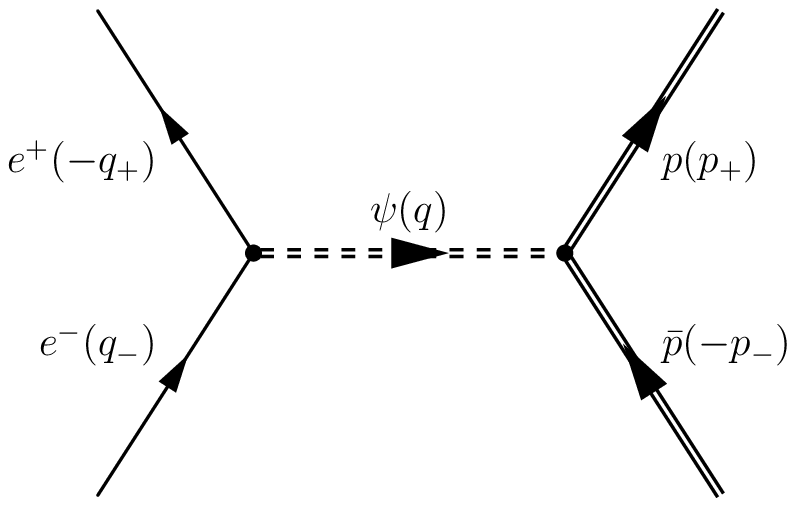}\label{fig.ChiDiagram}}
    }
    \caption{Feynman diagrams of processes $e^+ +e^-\to \bar p+p$ in Born approximation and with the quarkonium $\psi(3770)$ intermediate state.}
    \label{fig.fig1}
\end{figure}

We consider two mechanisms of creation of a $p\bar{p}$ in electron--positron collisions (see Fig.~\ref{fig.fig1})
\eq{
e^+(q_+)+e^-(q_-) \to p(p_+)+\bar{p}(p_-).
\label{eq.Process}
}
One proceeds through virtual photon intermediate state (see Fig.~\ref{fig.BornDiagram}), leading to the contribution to matrix element
\eq{
\M_B=\frac{4\pi\alpha}{s}G(s)J^e_\mu J^{p\mu},
}
where lepton $J^e_\mu$ and proton $J^p_\mu$ currents have a form:
\eq{
J^e_\mu=\bar{v}(q_+)\gamma_\mu u(q_-), \qquad J^p_\mu=\bar{u}(p_+)\gamma_\mu v(p_-), \nn
}
and $G(s)$ is the model--dependent proton formfactor. 

In the recent paper \cite{Ferroli:2010bi} the remarkable relation $F_1(\sqrt{s}\sim 2~\GeV)=1$, $F_2(\sqrt{s}\sim 2~\GeV)=0$
for proton form-factors near the threshold was obtained which meant, that proton in some environment near the $\sqrt{s}\sim 2-3~\GeV$
can be considered as a point-like particle. 
Assuming this facts and keeping in mind the closeness of the considered energy range to the $p\bar{p}$ threshold
we put further $G(s)=1$.
The  corresponding contribution to the differential cross section
\eq{
\frac{d\sigma}{d \Omega}=\frac{\alpha^2\beta}{4s}(2-\beta^2\sin^2\theta), \quad s=(q_++q_-)^2=4E^2,\quad \beta^2=1-\frac{m^2}{E^2},
}
where $m$ is the proton mass, $\sqrt{s}=2E$ is the total energy in center of mass reference frame (cmf), $E$ is the electron beam energy and the scattering angle $\theta$ is the cmf angle
between the 3-momenta of the initial electron $\vv{q_-}$ and the created proton $\vv{p_+}$.
The total cross section then
\eq{
\sigma_B(s)=\frac{2\pi\alpha^2 \beta (3-\beta^2)}{3s}.
}

\section{The quarkonium $\psi(3770)$ contribution: three gluon vertex}

The second mechanism (see Fig.~\ref{fig.ChiDiagram}) describes the conversion of electron--positron 
pair to $\psi(3770)$ with the subsequent conversion to the proton--antiproton pair through three 
gluon intermediate state (see Fig.~\ref{fig.PsiGGGPP}).

For this aim we put the whole matrix element as
\eq{
\M=\M_B+\M_\psi^{(3g)},
}
where the contribution with $\psi(3770)$ intermediate state is
\eq{
\M_\psi^{(3g)}=\frac{g_e}{s-M_\psi^2+iM_\psi\Gamma_\psi}J^e_\nu J_{(3g)}^\nu.
\label{eq.AmplitudeVia3G}
}
Here we assumed that vertex $\psi\to e^+ e^-$ has the same structure as $\gamma\to e^+ e^-$, i.e.:
\eq{
    J_{\psi\to e^+ e^-}^\mu = g_e \, J_e^\mu,
}
and the constant $g_e$ is defined via $\psi\to e^+ e^-$ decay ($g_e^2=12\pi\Gamma_{\psi\to e^+e^-}/M_\psi$) thus giving
$g_e = 1.6 \cdot 10^{-3}$ \cite{Beringer:1900zz}.

The current $J_{(3g)}^\nu$ which describes the transition of $\psi(3770)$ with momentum $q=2p$ into proton--antiproton
pair via three gluon intermediate state has the form (see Fig.~\ref{fig.PsiGGGPP}):
\begin{figure}
    \centering
    \mbox{
        \subfigure[]{\includegraphics[width=0.3\textwidth]{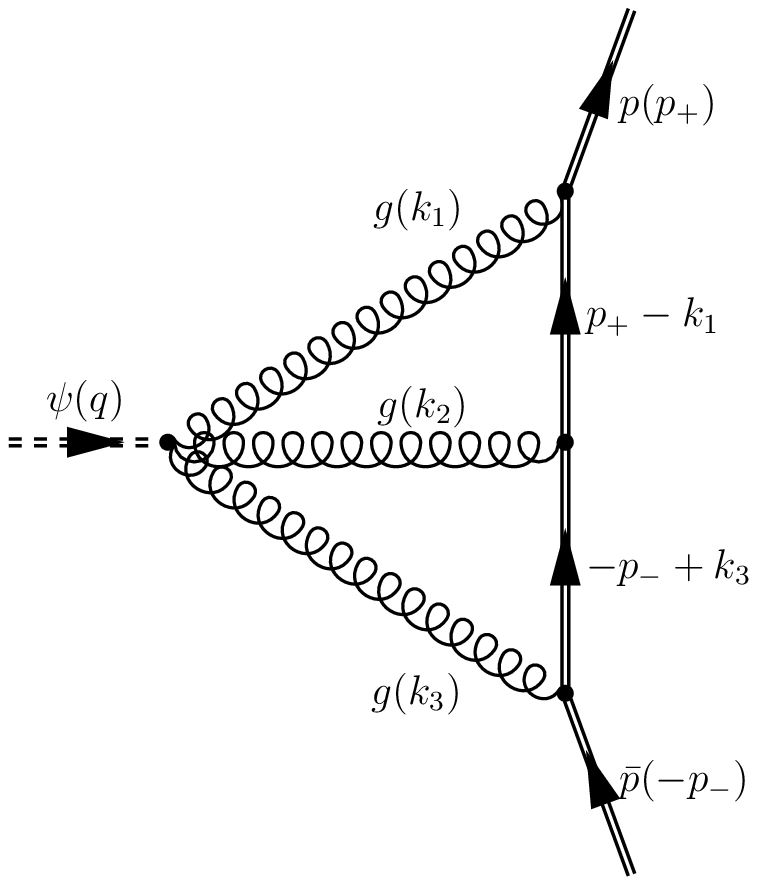}\label{fig.PsiGGGPP}}
        \quad\quad
        \subfigure[]{\includegraphics[width=0.3\textwidth]{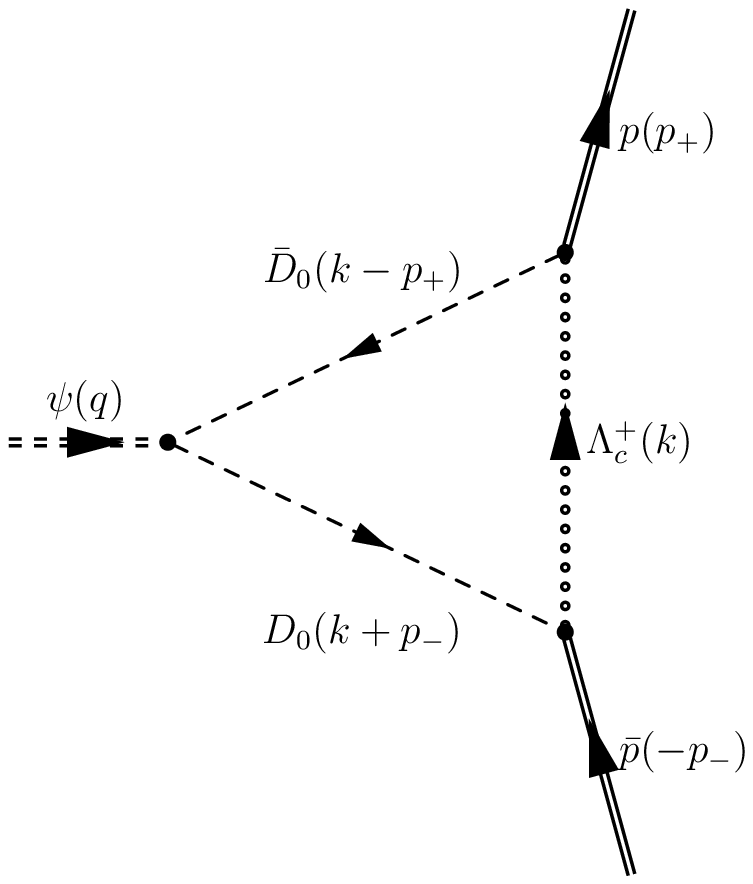}\label{fig.PsiDDPP}}
    }
    \caption{Vertexes of $\psi(3770)$ transition into proton--antiproton pair.}
    \label{fig.fig2}
\end{figure}
\eq{
    J_{(3g)}^\nu &=R \, \br{4\pi \alpha_s}^3 g_{col}\int\frac{d^4k_1 \, d^4k_2 \, d^4k_3 \, (2\pi)^{-8}}{k_1^2 \, k_2^2 \, k_3^2 \, ((p_+-k_1)^2-m^2)((p_--k_3)^2-m^2)}
    \times\nn\\
    &\times\delta(q-k_1-k_2-k_3)\brs{\bar{u}(p_+)\hat{O}^\nu v(p_-)},
    \label{eq.J3gDefinition}
}
where $\alpha_s$ is the strong interaction coupling which is associated with each gluon line and $\hat{O}^\nu$ is
\eq{
    \hat{O}_\nu&=\hat O_\nu^{\alpha\beta\gamma}\gamma_\alpha(\dd{p_+}-\dd{k_1}+m)\gamma_\beta(-\dd{p_-}+\dd{k_3}+m)\gamma_\gamma. \label{eq.hatQ}
}
and
\eq{
\hat O_\nu^{\alpha\beta\gamma}&=\Sp\brs{\hat{O}^{\alpha\beta\gamma}(\dd{p}+M)\gamma_\nu(\dd{p}-M)}, \nn \\
\hat{O}^{\alpha\beta\gamma}&= \frac{\gamma_\gamma(-\dd{p}+\dd{k_3}+M)\gamma_\beta(\dd{p}-\dd{k_1}+M)\gamma_\alpha}{\br{(p-k_3)^2-M^2}\br{(p-k_1)^2-M^2}} + \text{permutations},
}
where $p$ and $M$ are the 4-momentum and the mass of the charmed quark (anti-quark) inside $\psi(3770)$ state and one must take into account the contributions from all gluon lines permutations.
Color factor
\eq{
g_{col}=\bra{p} (3/4)d^{abc}t_at_bt_c \ket{p}=5/6
}
describes the interaction of gluons with quarks of the proton.
The quantity $R$ is connected with wave function of $\psi(3770)$ and is derived in \ref{appendix.PsiGGGVertex}.

Thus the contribution to the total cross section arising from the interference of relevant amplitudes has the form
\eq{
\delta\sigma_{3g}&=\frac{1}{8s} 2 \, \Re\brs{\sum_{\mbox{\tiny spins}} \int \M_B^*\, \M_\psi^{(3g)} \, d\Gamma_2},
}
where two--particle phase volume $d\Gamma_2$ is
\eq{
d\Gamma_2&=\frac{d^3p_+}{2E_+}\frac{d^3p_-}{2E_-}\frac{1}{4\pi^2}\delta^4(q-p_+-p_-)= \frac{\beta}{16\pi}d \cos\theta,
}
and $\theta$ is again the angle between the directions of initial electron $\vv{q_-}$ and the produced proton $\vv{p_+}$.

To perform the summation over spin states we use the method of invariant integration \cite{Byckling:1973}:
\eq{
    \sum_{\mbox{\tiny spins}} \int d\Gamma_2 \, J^{p *}_\mu J^{(3g)}_\nu
    =
    \frac{1}{3}\br{g_{\mu\nu}-\frac{q_\mu q_\nu}{q^2}}
    \int d\Gamma_2\sum_{\mbox{\tiny spins}} J^{p*}_\lambda J^{(3g)\lambda}=
-\frac{2s\beta}{3\pi}Q,
}
where
\eq{
Q=\frac{1}{4}\Sp\brs{(\dd{p_+} +m)\hat{O}_\lambda(\dd{p_-}+ m)\gamma_\lambda}.
}
Thus we get for the contribution to the total cross section
\eq{
\delta\sigma_{3g}&=\Re\br{\frac{S_{3g}\br{s}}{s-M_\psi^2+iM_\psi\Gamma_\psi}},
\label{eq.3gContribution}
}
where
\eq{
S_{3g}\br{s} &=-\frac{\alpha}{24} g_e \, g_{col} \, R \, \alpha_s^3 \, \beta \, Z\!\br{\beta},
\label{eq.S3g}\\
Z\br{\beta}&=\frac{4}{\pi^5 s}\int\frac{d^4k_1 \, d^4k_2 \, d^4k_3\, \delta(2p-k_1-k_2-k_3)}{k_1^2 \, k_2^2 \, k_3^2\br{(p_+ - k_1)^2-m^2}\br{(p_-k_2)^2-m^2}}
Q=
\nn\\
&=H(\beta)+i F(\beta), \label{eq.HandFdefinitions}
}
where $H(\beta)$ and $F(\beta)$ are correspondingly real and imaginary part of vertex $\psi \to 3g \to p\bar{p}$, i.e. function $Z\!\br{\beta}$.
Our approach consists in calculation of the $s$-channel discontinuity of $Z\!\br{\beta}$ with the subsequent restoration of real part $H(\beta)$ with the use of dispersion relation.
For this aim we use the Cutkosky rule for gluon propagators
\eq{
\frac{1}{(k_1^2+i0)}\frac{1}{(k_2^2+i0)}\frac{1}{(k_3^2+i0)} \to (-2\pi i)^3\delta(k_1^2)\delta(k_2^2)\delta(k_3^2).
}
This allows us integrate over phase volume of three gluon intermediate state as
\eq{
d\Phi_3&=\frac{d^4 k_1 d^4 k_2 d^4 k_3}{(2\pi)^5}\delta(k_1^2)\delta(k_2^2)\delta(k_3^2)\delta^4(2p-k_1-k_2-k_3)=  \nn \\
&=(2\pi)^{-5}\frac{1}{8}d x_1 d x_2 d\Omega_1 d\Omega_2\delta_c,
}
where
\ga{
\delta_c=\delta(c-p(x)), \qquad p(x)=1-2\frac{x_1+x_2-1}{x_1x_2}, \nn\\
x_i \equiv \frac{\omega_i}{E}, \qquad x_1+x_2+x_3=2, \nn
}
and $c$ is the cosine of the angle between directions $\vv{k_1}$ and $\vv{k_2}$. It is convenient to write the phase volume element in form
\eq{
d\Phi_3&=\frac{s\pi^2}{8(2\pi)^5}d x_1 \, d x_2 \, d\gamma \, \theta(1-x_1)\, \theta(1-x_2)\,\theta(x_1+x_2-1), \nn \\
d\gamma&=\frac{d \Omega_1 d \Omega_2}{4\pi^2}\delta_c=\frac{1}{\pi}\frac{d c_1 dc_2}{\sqrt{D}}, \quad D=1-c_1^2-c_2^2-p^2(x)+2c_1c_2p(x),
}
where $d \Omega_i$ is the phase volumes of the on mass shell gluons and $c_{1,2}\equiv\cos(\vv{p_+},\vv{k_{1,2}})$.
So we obtain $s$-channel discontinuity of $Z$ in the form:
\eq{
i\Delta_s Z=\int\limits_0^1 d x_1\int\limits_{1-x_1}^1 dx_2 \int d\gamma\frac{Q_1}{C_1C_2}=F(\beta),
\label{eq.DeltaSZ}
}
where $C_1=x_1(1-\beta c_1)$ and $C_2=x_2(1+\beta c_2)$ and the integration over
phase volume $d\gamma$ is performed in the kinematical region where $D>0$.
Explicit form of $i\Delta Z$ and $Q_1$ are given in \ref{appendix.AngularIntegrals}.
The angular integration can be performed using the form of the phase volume given above and the set of integrals given in \ref{appendix.AngularIntegrals}.

\begin{figure}
    \begin{center}\includegraphics[width=0.6\textwidth]{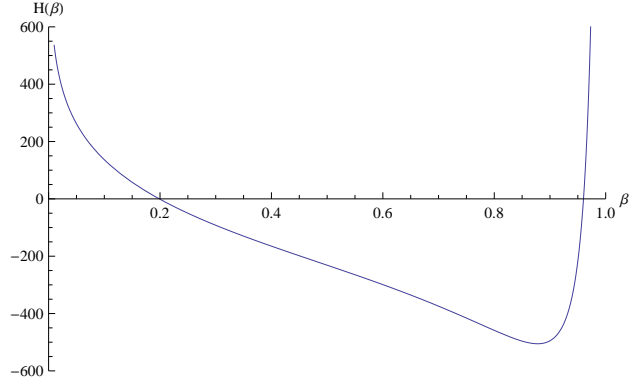}\end{center}
    \caption{\label{fig.H}
        The numerical estimation of the quantity $H(\beta)$ (see (\ref{eq.HandFdefinitions}) and (\ref{eq.H})) as a function of $\beta$.
    }
\end{figure}

As we are interested in the energy region close to the mass of resonance, we use some trick to restore the real part
of $Z$ by means of dispersion relations. For this aim we do a replacement
\eq{
Z(s) \to \Psi(s)=\frac{M_\psi^2}{s}Z(s), \qquad \Psi(s)=Z(s)\frac{M^2}{E^2}=\frac{M^2}{E^2}(H(\beta)+iF(\beta)).
}
We use the Cauchy theorem (un-substracted dispersion relation) to obtain the real part:
\eq{
H(\beta)&={\mathcal P}\frac{1}{\pi}\int\limits_0^1\frac{d\beta_1^2}{\beta_1^2-\beta^2}F(\beta_1)= \nn \\
&=\frac{1}{\pi}\brf{F(\beta)\ln\frac{1-\beta^2}{\beta^2}+\int\limits_0^1\frac{2\beta_1d \beta_1}{\beta_1^2-\beta^2}\brs{F(\beta_1)-F(\beta)}}. \label{eq.H}
}
The quantity $H(\beta)$ as a function of $\beta$ is shown in Fig.~\ref{fig.H}.
%

\section{The quarkonium $\psi(3770)$ contribution: $D^0$ mesons loop vertex}

It is known that the main contribution to the decay width of $\psi(3770)$ arise from the OZI non-violating
channels $\psi(3770) \to \bar{D}D$ \cite{Beringer:1900zz}.
However the contribution of $\bar{D}D$ state as an intermediate state converting to proton--antiproton
is expected to be small. The main reason for this is the absence of charmed quarks inside a proton.
In this section we will estimate the contribution of $D$ mesons loop to the process of our interest by using
only $D^0 \bar{D}^0$ loop in the vertex $\psi \to p\bar{p}$ (see Fig.~\ref{fig.PsiDDPP}).
The amplitude of the process (\ref{eq.Process}) with the $\psi$ intermediate state which converts via
$D^0 \bar{D}^0$ loop into proton--antiproton we write in the form similar to (\ref{eq.AmplitudeVia3G}):
\eq{
    \M_D =
    \frac{g_e}{s-M_\psi^2+iM_\psi\Gamma_\psi}
    J^e_\nu J_{D}^\nu.
}
where current $J_D^\mu$ has a form:
\eq{
    J_D^\mu &=
    \frac{g_{\psi DD}}{16 \pi^2}
    \int\frac{d^4k}{i \pi^2}\,
    g_D\!\br{\br{k-p_+}^2}
    g_D\!\br{\br{k+p_-}^2}
    \times\nn\\
    &\times
    \frac{\brs{\bar{u}\br{p_+} \gamma_5 \br{\dd{k}+M_{\Lambda^+_c}} \gamma_5 v\br{p_-}} \br{2k+p_--p_+}^\mu}
    { \br{k^2-M_{\Lambda^+_c}^2} \br{\br{k-p_+}^2-M_D^2} \br{\br{k+p_-}^2-M_D^2} },
    \label{eq.PsiDDPPCurrent}
}
where $g_{\psi DD}$ is the constant for vertex $\psi D^0 \bar{D}^0$ which can be estimated from
the decay width $\Gamma_{\psi \to D^0 \bar{D}^0} = 0.26~\keV$ \cite{Beringer:1900zz} which gives
\eq{
    g_{\psi DD} = \frac{4M_\psi \sqrt{3\pi \, \Gamma_{\psi \to D^0 \bar{D}^0}}}{\br{M_\psi^2-4M_D^2}^{3/4}} = 12.6.
}
The loop integral in (\ref{eq.PsiDDPPCurrent}) diverges in case of point-like particles.
Usually one uses some formfactor to cut this divergency \cite{Zhang:2009kr,Liu:2009dr}.
Following this tradition we use formfactors for the vertex $D^0 p \Lambda^+_c$ in the form
\cite{Reinders:1984sr}:
\eq{
    g_D\br{q^2} = \frac{2M_D^2 f_D}{m_u + m_c} \frac{g_{DN\Lambda}}{q^2 - M_D^2},
}
where $f_D \approx 180 - 200~\MeV$ and quark masses we choose as $m_u\approx 280~\MeV$ and $m_c = 1.27~\GeV$ \cite{Beringer:1900zz}.
The constant $g_{DN\Lambda}$ was estimated in \cite{Navarra:1998vi}:
\eq{
    g_{DN\Lambda} \approx 6.74.
}
Performing standard calculation of loop integral in (\ref{eq.PsiDDPPCurrent}) using Feynman trick to merge the denominators
one can write the contribution $D\bar{D}$ intermediate state to the cross section in the form similar to (\ref{eq.3gContribution}) as:
\eq{
\delta\sigma_{D}&=\Re\br{\frac{S_D\br{s}}{s-M_\psi^2+iM_\psi\Gamma_\psi}},
\label{eq.DDContribution}
}
where
\eq{
    S_D\br{s} &=
    \frac{\alpha}{24 \pi^2} g_e \, g_{DN\Lambda}^2 \, g_{\psi DD}
    \br{1+\frac{2m^2}{s}} \sqrt{1-\frac{4m^2}{s}} \, B_D\br{s},
    \label{eq.SD}
    \\
    B_D\br{s} &=
    \br{\frac{2M_D^2 f_D}{m_u + m_c}}^2
    \int\limits_0^1 dx \int\limits_0^{1-x} dy \, x y
    \times\nn\\
    &\times
    \brf{
        \frac{1}{\br{d\br{s}+i \epsilon}^2}
        +
        \frac{2m x}{\br{d\br{s}+i \epsilon}^3}
        \frac{s-4m^2}{s+2m^2}
        \br{M_{\Lambda^+_c} - m\br{1-x}}
    },
    \\
    d\br{s} &= M_{\Lambda^+_c}^2 x + M_D^2 \br{1-x} - m^2 x \br{1-x} - s y \br{1-x-y}.
}

\section{Discussion}

\begin{figure}
    \begin{center}\includegraphics[width=0.6\textwidth]{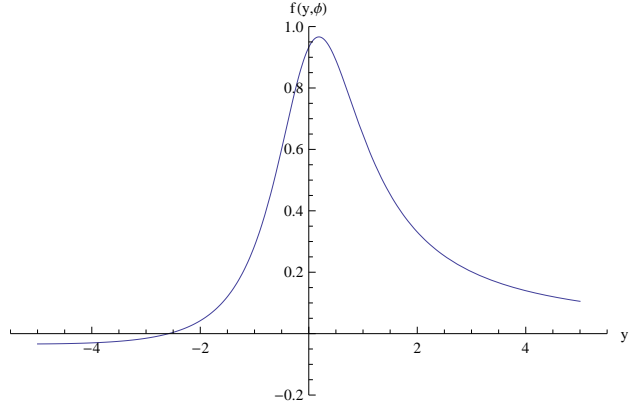}\end{center}
    \caption{\label{fig.f}
        The numerical estimation of the quantity $f(y,\phi)$ (see (\ref{eq.f})) as a function of $\beta$.
    }
\end{figure}

In order to see the relative contribution of different mechanisms to the phase we will consider
first the contribution of three gluons in the intermediate state.
The total cross section then has a form
\eq{
\sigma(s)&=\sigma_B(s) + \delta\sigma_{3g}(s), \qquad
\frac{\delta\sigma_{3g}(s)}{\sigma_B(s)}=B(\beta)f(y,\phi),
\label{eq.total1}
}
where
\eq{
B(\beta)&=\frac{g_eg_{col}R\alpha_s^3}{32\alpha(3-\beta^2)}\frac{PM_\psi}{\Gamma_\psi}, \nn \\
f(y,\phi)&=\frac{y\cos\phi+\sin\phi}{y^2+1}, \qquad y=\frac{s-M_\psi^2}{M_\psi\Gamma_\psi}, \label{eq.f}
}
%
and the quantities $P$ and $\phi$ are defined as
\ga{
H+iF = Pe^{i \phi}, \qquad P=\sqrt{H^2+F^2}; \\
R=\frac{1}{9}\sqrt{\frac{2}{\pi}}\alpha_s^{3/2}; \qquad g_e=\sqrt{\frac{12\pi\Gamma_{ee}}{M_\psi}}; \qquad g_{col}=\frac{5}{6}\nn.
}
The function $f(y,\phi)$ is shown in Fig.~\ref{fig.f}.
%
%
At the point of $\psi(3770)$ resonance, $\beta=\beta_0=0.86$,  we have both quantities $F$ and $H$ negative and thus
the phase $\phi$ is equal to
\eq{
    \phi =  \arctan\br{\frac{F(\beta_0)}{H(\beta_0)}} + 180^o = 67^o + 180^o = 247^o.
    \label{eq.Result1}
}
%
The ratio of the $B(\beta_0)$ to $P$ is
\eq{
	B(\beta_0)=3 \cdot 10^{-5} P,\qquad P=1396.
    \label{eq.Value}
}

It is known that the main contribution to the width of $\psi(3770)$ arise from the OZI non-violating
channels $\psi(3770) \to \bar{D}D$ \cite{Beringer:1900zz}. However the contribution of $\bar{D}D$ state as an intermediate state
converting to proton--antiproton  is small. Main reason of it is the absence of charm quarks inside a proton.
In order to demonstrate this we add the $D$-loop contribution $\delta \sigma_D$ from (\ref{eq.DDContribution})
to the cross section in (\ref{eq.total1}), i.e.:
\eq{
    \sigma(s)&=\sigma_B(s) + \delta\sigma_{3g}(s) + \delta \sigma_D(s),
}
and then, to calculate the phase $\phi$, we need to use complete expressions for the amplitudes, i.e. $S_{3g}\br{s}$ from (\ref{eq.S3g}) and $S_D\br{s}$ from (\ref{eq.SD}).
This gives the following result for the phase:
\eq{
    \phi &= \arctan\br{\frac{\Im\br{S_{3g}\br{M_\psi^2}} + \Im\br{S_D\br{M_\psi^2}}}{\Re\br{S_{3g}\br{M_\psi^2}} + \Re\br{S_D\br{M_\psi^2}}}} + 180^o =
    \nn\\
    &= 81^o + 180^o = 261^o,
    \label{eq.Result2}
}
and thus we conclude that $D$-meson loop contribution to the phase is rather small and the main contribution to the
phase goes from three gluon intermediate state.

We should also notice that we did not evaluate the contribution of a square of amplitude with $\psi(3770)$ intermediate state.
It is small compared
with the contribution of interference of Born amplitude with the one with $\psi(3770)$ meson and will be estimated elsewhere.
It does not exceed ten percents.

The quantities for phase $\phi$ in (\ref{eq.Result1}) and in (\ref{eq.Result2})
are in good agreement with recent experimental data for phase at BES III collaboration \cite{Ablikim:2014jrz}.

\section*{Acknowledgements}

The authors acknowledge to RFBR grant no. 11-02-00112-а for financial support.
One of us (Yu.~M.~B.) also acknowledges JINR grant No. 13-302-04 of 2013 year for support.
This work was supported by National Natural Science Foundation of China
under Contract No. 11175187.

\appendix

\section{Vertex $\psi \to 3g$}
\label{appendix.PsiGGGVertex}

To restore the quantity $R$ from (\ref{eq.J3gDefinition}) we calculate the width of $\psi(3770)$ resonance decay
into three gluons.
Let us consider the conversion of the bound state with quantum numbers $J^{PC}=1^{--}$ to three real massless
vector bosons.
Similar problem was solved years ago for the problem of ortho-positronium decay \cite{Pomeranchuk:1948,Ore:1949te}
For the case of ortho-positronium $Ops$ decay, we start from matrix element
of the process:
\eq{
    Ops \to \gamma(k_1) + \gamma(k_2) + \gamma(k_3),
}
which has the form:
\eq{
\M_{Ops}=A\frac{1}{m_e^4}O_\sigma^{\mu\nu\lambda}e_\mu(k_1)e_\nu(k_2)e_\lambda(k_3)\epsilon_\sigma(q),
\label{eq.OpsAmplitude}
}
with $e(k_i)$ and $\epsilon(q)$ are the polarization vectors of photons and the ortho-opositronium respectively. The quantity $A$ includes the information
on the wave function of ortho-positronium. Operator
\eq{
&O_\sigma^{\mu\nu\lambda}e_\mu(k_1)e_\nu(k_2)e_\lambda(k_3) = \frac{1}{4} \Sp\brs{\hat{Q} (\dd{p} +m_e)\gamma_{\sigma} (\dd{p} -m_e)};
\\
\hat{Q}&=\frac{1}{x_1 x_3}\dd{e_3}(-\dd{p}+\dd{k_3} +m_e)\dd{e_2} (\dd{p} -\dd{k_1} + m_e)\dd{e_1} + \text{cyclic permutations},
}
describes the electron loop. Using the amplitude (\ref{eq.OpsAmplitude}) we obtain for the decay width
\eq{
\Gamma_{Ops} &= \frac{1}{12 m_e}\int\sum_{spins}|\M_{Ops}|^2 \cdot \frac{m_e^2\pi^2}{(2\pi)^5}d^2x \cdot \frac{(4\pi\alpha)^3}{3!}=
\nn\\&=
\frac{64}{9}m_e A^2(\pi^2-9)\alpha^3.
}
Comparing this value with the known result $\Gamma_{Ops}=(2 m_e/(9\pi))(\pi^2-9)\alpha^6$  we conclude that
\eq{
A=\frac{\alpha^{3/2}}{4\sqrt{2\pi}}.
}
Here we used the following formulae
\ga{
\sum_{spins}\frac{1}{m_e^8}\brm{O_\sigma^{\mu\nu\lambda}e_\mu(k_1)e_\nu(k_2)e_\lambda(k_3)}^2=256 Q(x), \nn \\
Q(x)=\frac{1}{(x_1x_2x_3)^2}[x_1^2(1-x_1)^2+x_2^2(1-x_2)^2+x_3^2(1-x_3)^2],  \nn \\
\int d^3x \, \delta(2-x_1-x_2-x_3) \, Q(x)=\pi^2-9.\nn
}
\begin{figure}
    \centering
    \includegraphics[width=0.4\textwidth]{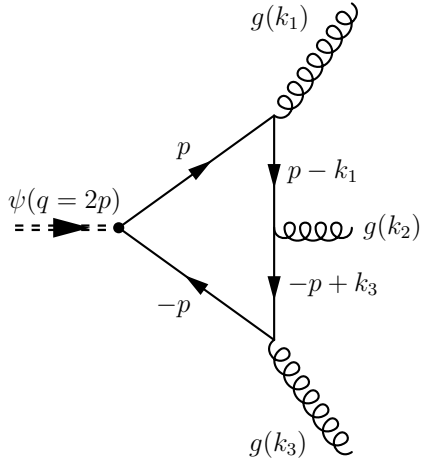}
    \caption{The diagram of $\psi(3770)$ decay into three gluons.}
    \label{fig.PsiGGG}
\end{figure}
For the case of decay of $\psi(3770)$ to three gluons with the subsequent turning them to hadrons we define the amplitude in the form similar to (\ref{eq.OpsAmplitude})
(see Fig.~\ref{fig.PsiGGG}):
\eq{
\M_{\psi \to 3g} = R(4\pi\alpha_s)^{3/2} \frac{1}{4}d^{abc} \, e_{\mu}^a(k_1)e_{\nu}^b(k_2)e_{\lambda}^c(k_3) \frac{1}{M^4}O_{\sigma}^{\mu\nu\lambda}\varepsilon^{\sigma}(q),
}
with $q=2p$ and $\varepsilon(q)$ are the momentum and the polarization vector of $\psi(3770)$.
The decay width then reads as:
\eq{
\Gamma_{\psi \to 3g} = \frac{80}{27}M_{\psi} R^2(\pi^2-9)\alpha_s^3.
}
And comparing this result with the known one \cite{Appelquist:1974zd,Kwong:1987ak}:
\eq{
    \Gamma_{\psi \to 3g}=\frac{160 M_{\psi}}{2187\pi}(\pi^2-9)\alpha_s^6,
}
we conclude that
\eq{
    R=\frac{1}{9}\sqrt{\frac{2}{\pi}}\alpha_s^{3/2} \approx 0.0146,
}
if one assumes that $\alpha_s \approx 0.3$.
Note that both $A$ and $R$ are real.

\section{Angular integrals}
\label{appendix.AngularIntegrals}

\begin{figure}
    \begin{center}\includegraphics[width=0.6\textwidth]{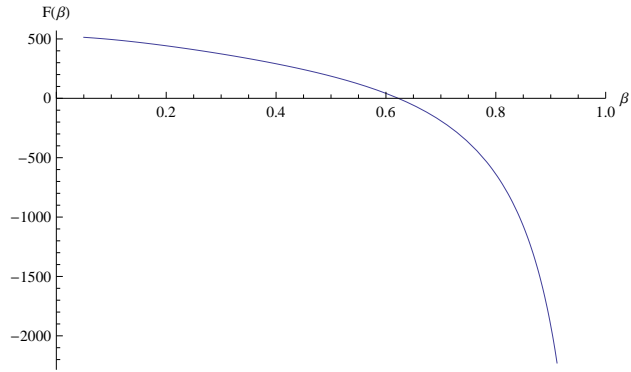}\end{center}
    \caption{\label{fig.F}
        The numerical estimation of the quantity $F(\beta)$ (see (\ref{eq.F})) as a function of $\beta$.
    }
\end{figure}

In this section we present the angular integrals which are relevant for the integration in (\ref{eq.DeltaSZ}):
\eq{
&\frac{1}{\pi}\int \frac{dc_1 dc_2}{C_1 C_2 \sqrt {D}}\biggl\{1; \,\,C_1;\,\, C_2;\,\, C_1C_2;\,\, C_1^2;\,\, C_2^2;\,\, C_2^3;\,\, C_1^2C_2;\,\, C_2^2C_1\biggr\} = \nn \\
&\qquad =\biggl\{J_{00},J_{10},J_{01},J_{11},J_{20},J_{02},J_{03},J_{21},J_{12}\biggr\},
}
where
\ga{
J_{00}=\frac{1}{x_1 x_2}I(x),\qquad J_{10}=\frac{1}{x_2}L,\qquad J_{01}=\frac{1}{x_1}L,\qquad J_{11}=2, \nn \\
J_{20}=\frac{x_1}{x_2}[L(1+p) - 2p], \qquad J_{02}=\frac{x_2}{x_1}[L(1+p) - 2p], \nn \\
J_{21}=2x_1,\qquad J_{12}= 2x_2, \nn\\
J_{03}=\frac{x_2^2}{2x_1}[(1+\beta^2+4p +(3-\beta^2)p^2)L +2(1-4p-3p^2)], \nn
}
and
\ga{
p = p(x)=1-\frac{2}{x_1 x_2}(x_1 +x_2 -1);\nn\\
C_1 = x_1(1-\beta c_1);\qquad C_2 = x_2(1+\beta c_2),\nn \\
I(x)=\frac{2}{\sqrt{d}}\ln\frac{1+\beta^2p(x)+\sqrt{d}}{1-\beta^2}, \nn\\
d=(1+\beta^2p(x))^2-(1-\beta^2)(1-p(x)^2), \qquad L=\frac{1}{\beta}\ln\frac{1+\beta}{1-\beta}. \nn
}
The explicit expression for $Q_1=T^{\alpha\beta\gamma}_\lambda R^{\alpha\beta\gamma}_\lambda$ from (\ref{eq.DeltaSZ}) is
 \eq{
T^{\alpha\beta\gamma}_\lambda &=\frac{1}{4}\Sp\brs{\hat{O}^{\alpha\beta\gamma}(\dd{p}+M)\gamma_\lambda (\dd{p}-M)}; \nn \\
R^{\alpha\beta\gamma}_\lambda &=\frac{1}{4}\Sp\brs{(\dd{p_-}-m)\gamma_\lambda (\dd{p_+}+m)\gamma_\alpha (\dd{p_+}-\dd{k_1}+m)\gamma_\beta(-\dd{p_-}+\dd{k_2}+m)\gamma_\gamma}, \nn
 }
where
\eq{
 \hat{O}^{\alpha\beta\gamma}&=
 \frac{1}{x_1}\brs{
    \frac{1}{x_2}\gamma_\beta(-\dd{p}+\dd{k_2}+M)\gamma_\gamma+
    \frac{1}{x_3}\gamma_\gamma(-\dd{p}+\dd{k_3}+M)\gamma_\beta
 }(\dd{p}-\dd{k_1}+M)\gamma_\alpha+ \nn \\
 &+\frac{1}{x_2}\brs{
    \frac{1}{x_3}\gamma_\gamma(-\dd{p}+\dd{k_3}+M)\gamma_\alpha+
    \frac{1}{x_1}\gamma_\alpha(-\dd{p}+\dd{k_1}+M)\gamma_\gamma
 }(\dd{p}-\dd{k_2}+M)\gamma_\beta+ \nn \\
 &\frac{1}{x_3}\brs{
    \frac{1}{x_2}\gamma_\beta(-\dd{p}+\dd{k_2}+M)\gamma_\alpha+
    \frac{1}{x_1}\gamma_\alpha(-\dd{p}+\dd{k_1}+M)\gamma_\beta
 }(\dd{p}-\dd{k_3}+M)\gamma_\gamma.\nn
}
After calculation of traces and simplifications one gets
\eq{
Q_1&=\frac{32}{x_1x_2x_3}
\left\{P_{00}+C_1P_{10}+C_2P_{01}+C_1C_2P_{11}+C_1^2P_{20}+\right.\nn\\
&+\left.C_2^2P_{02}+C_2^3P_{03}+
C_1^2C_2P_{21}+C_1C_2^2P_{12}\right\},
}
where coefficients $P_{ij}$ have a form
\eq{
P_{00}&=-4(-1+x_1)^2(-4+x_1+b(-1+b+x_1))- \nn \\
&-4(17+b^2(-2+x_1)-26x_1+8x_1^2+b(-1+x_1)(1+2x_1))x_2-\nn\\
&-4(-17+9x_1+b(4+b+x_1))x_2^2+8(-2+b)x_2^3; \nn \\
P_{10}&=4(-1+x_1)^2-2(8+x_1(-11-3b+6x_1))x_2 +4(1+b-5x_1)x_2^2-4x_2^3, \nn \\
P_{01}&=2(10-8x_1-3x_1^2+2x_1^3+2(-2+x_1)(4+3x_1)x_2 +2(3+x_1)x_2^2 +\nn\\
&+
b[x_1^2-2(-2+x_2)(-1+x_2)+2x_1(1+x_2)]), \nn \\
P_{11}&=2(-4+x_1+x_1^2+(5+x_1)x_2-2b(-1+x_1+x_2)),\nn \\
P_{20}&=2x_2(-2-b+3x_1+3x_2), \nn \\
P_{02}&=-2(2(-1+x_2)+x_1(b+2(-1+x_1+x_2))), \nn \\
P_{21}&=2(2-x_1-x_2), \nn \\
P_{12}&=2 x_1, \nn \\
P_{03}&=2 x_1, \nn
}
here we use the notation that $b=\beta^2$.
The contribution to imaginary part $F(\beta)$ then reads as
\eq{
F(\beta)&=\int\limits_0^1 d x_1\int\limits_{1-x_1}^1 d x_2\frac{32}{x_1x_2x_3}
\left\{P_{00}J_{00}+P_{10}J_{10}+P_{01}J_{01}+\right.\nn\\
&+\left.
P_{11}J_{11}+P_{20}J_{20}+P_{02}J_{02}+P_{03}J_{03}+P_{21}J_{21}+P_{12}J_{12}\right\}.
\label{eq.F}
}
Numerically the quantity $F(\beta)$ as a function of $\beta$ is presented in Fig.~\ref{fig.F}.










\end{document}